\documentstyle[12pt]{ioplppt}
\eqnobysec
\begin{document}
\jl{1}
\title{On the Finite--Temperature Generalization of \\ the C-theorem 
and the Interplay  between \\ Classical and Quantum Fluctuations}[On the
Generalization of the C-theorem]
\author{Daniel M. Danchev$^{*}$ and Nicholay S. Tonchev$^{**}$}
\address{$^{*}$Institute of Mechanics, Bulgarian Academy of Sciences, Acad. G.\\
Bonchev St. bl. 4, 1113 Sofia, Bulgaria\\
$^{**}$Istitute of Solid State Physics, Bulgarian Academy of Sciences,\\
Tzarigradsko chauss\'{e}e blvd. 72, 1784 Sofia, Bulgaria}
\date{March 10, 1999}

\begin{abstract}
The behavior of the finite-temperature $C$-function, defined by Neto and
Fradkin [Nucl. Phys. B {\bf 400}, 525 (1993)], is analyzed within a $d$%
-dimensional exactly solvable lattice model, recently proposed by Vojta
[Phys. Rev. B {\bf 53}, 710 (1996)], which is of the same universality class
as the quantum nonlinear O(n) sigma model in the limit $n\rightarrow \infty
. $ The scaling functions of $C$ for the cases $d=1$ (absence of long-range
order), $d=2$ (existence of a quantum critical point), $d=4$ (existence of a
line of finite temperature critical points that ends up with a quantum
critical point) are derived and analyzed. The locations of regions where $C$
is monotonically increasing (which depend significantly on $d$) are exactly
determined. The results are interpreted within the finite-size scaling
theory that has to be modified for $d=4.$
\end{abstract}


\maketitle

\section{INTRODUCTION}

The original Zamolodchikov's $C$-theorem is related to zero-temperature
systems. It establishes the existence of a dimensionless function $C$ of the
coupling constants with monotonic properties along the renormalization group
trajectories \cite{Z86}. The assumptions presented in the proof are related
with the energy-momentum conservation, the rotational and translational
symmetries, and positivity in a two dimensional quantum field theory. The
behavior of the $C$-function reflects the role the quantum fluctuations and
it is useful in determining the qualitative futures of the theory away from
the criticality. At the fixed points it takes the value of the central
charge of the corresponding conformal field theory. Since the basic
assumptions underlying the $C$-theorem are not specific to two dimensions
only, a considerable interest exists in generalization of the
Zamolodchikov's result for dimensionalities different from two as well as
for nonzero temperatures \cite{J91} - \cite{PV98}. Earlier efforts (see \cite
{J91} and references therein) have been devoted to finding a version of the $%
C$-theorem valid in four dimensions. There the approach was based on a
careful investigation of the form of the trace of the energy-momentum
tensor, written in terms of finite local composite operators. Despite of
being able to write expressions for the Zamolodchikov's equations for the $C$%
-function, similar to the case of two dimensions, it turns out that it is
not possible to demonstrate the monotonicity property. Let us note also the
fact that the $3$-d analog of central charge \cite{Cardy87} is not
equivalent to the universal number characterizing the size dependence of the
free energy at the critical point \cite{S93} (which is always the case in $2$%
-d conformal field theory). This fact indicates that a straightforward
generalization of the Zamolodchikov's $C$-theorem is not to be expected for
a general $d$. See also Ref. \cite{Z97}, where different approaches to the
problem have been offered and where it was shown that no direct relations
exist between the ``finite temperature $C$-theorem'' and the Zamolodchikov's 
$C$-theorem at zero temperature. In the present study an approach, proposed
by Netto and Fradkin \cite{CN93} (in some sense a thermodynamic one), for
finding a candidate for $C$-function will be considered. In \cite{CN93} the
following dimensionless function is defined 
\begin{equation}
C\left( \beta ,g,a|d\right) =-\beta ^{d+1}\frac{v^d}{n(d)}\left[ f\left(
\beta ,g,a|d\right) -f\left( \infty ,g,a|d\right) \right] ,  \label{Cdef}
\end{equation}
where $\beta $ is the inverse temperature ($\beta =1/T$) with the Boltzman
constant $k_B=1$, $g$ is a set of dimensionful coupling constants, $f\left(
\infty ,g,a|d\right) \equiv E_0\left( g,a|d\right) $ is the zero-temperature
energy density, i.e. the energy of the ``infinite'' in the inverse
temperature system, $f\left( \beta ,g,a|d\right) $ is the full free energy
density (per unit volume) of the system, and $a$ is the characteristic
length scale of the lattice. Here $n(d)$ is a positive real number (which
depends only on the dimensionality $d$ of the system) and $v$ is the
characteristic velocity (e.g. the velocity of quasiparticles) in the system.
The function $C\left( \beta ,g,a|d\right) $ is considered to be the $d$%
-dimensional nonzero temperature extension of the Zamolodchikov's
C-function. It is supposed to be {\it positive}, and, in the regions where
the quantum fluctuations dominate, a {\it monotonically increasing function
of the temperature}. In \cite{CN93} the numbers $n(d)=\Gamma ((d+1)/2)\zeta
(d+1)/\pi ^{(d+1)/2}$ for bosons ($\zeta (x)$ is the Riemann dzeta function, 
$\Gamma (x)$ is the gamma function) and $n(d)=\Gamma ((d+1)/2)\zeta
(d+1)(2-2^{1-d})/\pi ^{(d+1)/2}$ for fermions have been suggested.
Obviously, the exact choice of $n(d)$ does not effect the monotonicity
properties of the $C$-function. Functions analogous to the one defined in (%
\ref{Cdef}) have been discussed also in Refs. \cite{S93} - \cite{PV98}.

For a general $d$ the existence of phase transitions in the system, as well
as the interplay between the quantum and classical fluctuations, makes the
analysis of the behavior of the $C$-function difficult from a general point
of view. That is why, for any $d$ $\neq 1$ the properties of $C$ have been
considered on the examples of concrete models: the free massive filed
theories (for any $d$) \cite{CN93}, the Ising model in a transverse field
(for $d=1$) \cite{CN93} and the quantum nonlinear sigma model (QNL$\sigma $%
M) in the limit $N\rightarrow \infty $ and $d=2$ \cite{CN93}, \cite{S93}.
(Recently, the value of the $C$-function at the critical point as a function
of $d$ , $1<d<3$, has been calculated in \cite{PV98} for that model).

In the present article we will consider the $d$-dependence of the
monotonicity property of the $C$-function within the framework of an exactly
solvable lattice model. We will explicitly demonstrate the crucial role that
the existence of a quantum ($T=0$) and/or classical ($T\neq 0$) critical
points plays for the behavior of $C$ in different regions of the phase
diagram in the plane temperature - parameter controlling the quantum
fluctuations.

The article is organized as follows. In Section II we briefly describe the
model and in Section III present the basic exact analytical expressions for
the free energy of the bulk system at nonzero temperature. Section IV
contains the analysis of the behavior of the $C$-function in dimensions $1,2$
and $4$ in different regimes of the parametric space of the model. Section V
presents a finite-size scaling interpretation of the results. The article
closes with concluding remarks given in Section VI.

\section{The Model}

\label{model} The model we consider describes a magnetic ordering due to the
interaction of quantum spins. The Hamiltonian has the following form \cite
{V96} 
\begin{equation}
{\cal H}=\frac 12g\sum_\ell {\cal P}_\ell ^2-\frac 12\sum_{\ell \ell
^{\prime }}{J}_{\ell \ell ^{\prime }}{\cal S}_\ell {\cal S}_{\ell ^{\prime
}}+\frac \mu 2\sum_\ell {\cal S}_\ell ^2,  \label{model1}
\end{equation}
where ${\cal S}_\ell $ are spin operators at site $\ell $, the operators $%
{\cal P}_\ell $ are ``conjugated'' momenta (i.e. $[{\cal S}_\ell ,{\cal S}%
_{\ell ^{\prime }}]=0$, $[{\cal P}_\ell ,{\cal P}_{\ell ^{\prime }}]=0$, and 
$[{\cal P}_\ell ,{\cal S}_{\ell ^{\prime }}]=i\delta _{\ell \ell ^{\prime }}$%
, with $\hbar =1$), the coupling constants ${J}_{\ell ,\ell ^{\prime }}={J}$
are between nearest neighbors only,~ the coupling constant $g$ is introduced
so as to measure the strength of the quantum fluctuations (below it will be
called quantum parameter), and, finally, the spherical field $\mu $ is
introduced to insure the fulfillment of the constraint $\left\langle {\cal S}%
_\ell ^2\right\rangle =1$ . Here $\left\langle \cdots \right\rangle $
denotes the standard thermodynamic average taken with ${\cal H}$.

In the thermodynamic limit the reduced free energy $\tilde{f}_\infty \left(
\beta ,g|d\right) =$ $f_\infty \left( \beta ,g|d\right) /\sqrt{gJ}$ takes
the form \cite{rem2} 
\begin{eqnarray}
\lambda \tilde{f}_\infty \left( t,\lambda |d\right)& = &
\sup_\phi \left\{ \frac %
t{\left( 2\pi \right) ^d}\int_{-\pi }^\pi dq_1\ldots \int_{-\pi }^\pi
dq_d   \right.   \\
& \times & \left. \ln \left[ 2\sinh \left( \frac \lambda {2t}\sqrt{\phi
+2\sum_{i=1}^d(1-\cos q_i)}\right) \right] -\frac 12\phi \right\} -d,
\nonumber
\label{be}
\end{eqnarray}
where we have introduced the notations: $\lambda =\sqrt{g/J}$ is the
normalized quantum parameter, $t=\frac TJ$ is the normalized temperature and 
$\phi =\frac \mu J-2d$ is the shifted spherical field. The supremum is
attained at a solution of the mean--spherical constraint that reads 
\begin{equation}
1=\frac t{\left( 2\pi \right) ^d}\sum_{m=-\infty }^\infty \int_{-\pi }^\pi
dq_1\ldots \int_{-\pi }^\pi dq_d\frac 1{\phi +2\sum_{i=1}^d(1-\cos
q_i)+b^2m^2},  \label{sfe}
\end{equation}
where $b=2\pi t/\lambda .$

Eqs.~(\ref{be}) and (\ref{sfe}) provide the basis for studying the critical
behavior of the model under consideration.

The critical behavior and some finite size properties of this model have
been considered in \cite{HPT98}, \cite{HDPT97} for $1<d<3$. Below we present
a brief sketch of the derivation of the bulk free energy for $d=1,2,4$ at
low temperatures.

\section{The free energy at low temperatures}

\label{FEN}By using the identities: 
\begin{equation}
\ln \frac{\sinh b}{\sinh a}=\frac 12\sum_{m=-\infty }^\infty \ln \frac{%
b^2+\pi ^2m^2}{a^2+\pi ^2m^2},  \label{id1}
\end{equation}
where $ab>0$, $a,b$ are arbitrary real numbers, 
\begin{equation}
\ln \left( a+b\right) =\ln a+\int_0^\infty \exp \left( -ax\right) \left(
1-\exp \left( -bx\right) \right) \frac{dx}x,  \label{id2}
\end{equation}
where $a>0,$ $a+b>0,$ and the Jacobi identity after some algebra at low
temperatures ($\frac \lambda t\gg 1$) the expression for the free energy (%
\ref{be}) can be rewritten in the form 
\begin{equation}
2\lambda \tilde{f}_\infty \left( t,\lambda |d\right) =\lambda a(\phi
,d)-(\phi +2d)-\lambda s(\phi ,b,d),  \label{gfe}
\end{equation}
where 
\begin{equation}
a(\phi ,d)=\frac 1{2\sqrt{\pi }}\int_0^\infty \frac{dx}{x^{3/2}}\exp \left(
-x\phi \right) \left[ 1-\left( \exp \left( -2x\right) I_0(2x)\right)
^d\right] +\sqrt{\phi },  \label{fd}
\end{equation}
\begin{equation}
s(\phi ,b,d)=2\int_0^\infty \frac{dx}x\left( 4\pi x\right) ^{-(d+1)/2}\exp
\left( -x\phi \right) R\left( \frac{\pi ^2}{xb^2}\right) ,  \label{ftt}
\end{equation}
\begin{equation}
R(x)=\sum_{m=1}^\infty \exp \left( -xm^2\right) ,  \label{R}
\end{equation}
$I_0(x)$ is a modified Bessel function, and $\phi $ in (\ref{gfe}) is the
solution of the corresponding spherical field equation~that follows from (%
\ref{gfe}) by requiring that the first partial derivative of the right hand
side of (\ref{gfe}) with respect to $\phi $ is zero. The above expressions
are valid for {\it any} $d.$

In the remainder we will consider the dimensions $d=1,d=2$ and $d=4.$ Then
it can be shown that:

a) for $d=1$

\begin{eqnarray}
a(\phi ,1) &=&\frac 12\;_2F_1\left( -\frac 14,\frac 14,1,\frac 4{(2+\phi )^2}%
\right) \sqrt{2+\phi },  \label{a1} \\
s(\phi ,b,1) &=&-\frac b{2\pi ^2}\sqrt{\phi }\sum_{m=1}^\infty
m^{-1}K_1\left( \frac{2\pi m\sqrt{\phi }}b\right)  \label{s1}
\end{eqnarray}
where $_2F_1$is the hypergeometric function and $K_1(x)$ is the MacDonald
function (second modified Bessel function).

b) for $d=2,$ and $\phi <<1$

\begin{eqnarray}
a(\phi ,2) &\simeq &a(0,2)+{\cal W}_2(0)\phi -\frac 1{6\pi }\phi ^{3/2}
\label{a2} \\
s(\phi ,b,2) =&-& \left( \frac b{2\pi }\right) ^3\left[ \frac{\sqrt{\phi }}b%
{\rm Li}_2\left( \exp \left( -\frac{2\pi \sqrt{\phi }}b\right) \right)
\right.
\nonumber \\
&+& \left.
\frac 1{2\pi }{\rm Li}_3\left( \exp \left( -\frac{2\pi \sqrt{\phi }}b\right)
\right) \right]  \label{s2}
\end{eqnarray}
where 
\begin{equation}
{\cal W}_d(\phi )=\frac 1{2(2\pi )^d}\int_{-\pi }^\pi dq_1\ldots \int_{-\pi
}^\pi dq_d\left( \phi +2\sum_{i=1}^d\left( 1-\cos q_i\right) \right) ^{-1/2}.
\label{bulk4}
\end{equation}
is a Watson type integral, ${\cal W}_2(0)\approx $ $0.3214$, and ${\rm Li}%
_n(x)$ is the polylogarithmic function.

c) for $d=4,$ and $\phi <<1$

\begin{equation}
a(\phi ,4) \simeq a(0,4)+{\cal W}_4(0)\phi -\frac 12r\phi ^2+\frac 1{30\pi
^{3/2}}\phi ^{5/2},  \label{a4}
\end{equation}
\begin{eqnarray}
\lefteqn{
s(\phi ,b,4) =-\left( \frac b{2\pi }\right) ^5\left[ \frac \phi {b^2}{\rm %
Li}_3\left( \exp \left( -\frac{2\pi \sqrt{\phi }}b\right) \right) \right.}
\nonumber \\
& & \left. + \frac{3%
\sqrt{\phi }}{2\pi b}{\rm Li}_4\left( \exp \left( -\frac{2\pi \sqrt{\phi }}b%
\right) \right) +\frac 3{4\pi ^2}{\rm Li}_5\left( \exp \left( -\frac{2\pi 
\sqrt{\phi }}b\right) \right) \right]  \label{s4}
\end{eqnarray}
where ${\cal W}_4(0)\approx $ $0.1891$ and 
\begin{equation}
r=\int_0^\infty \sqrt{x}\left[ \exp \left( -2x\right) I_0\left( 2x\right)
\right] ^4dx\approx 0.0677.  \label{r}
\end{equation}
Now we have the basic expressions needed to analyze the behavior of the
finite-temperature $C$-function as defined by Eq. (\ref{Cdef}).

\section{The behavior of the $C$-function}

\subsection{The case $d=1$}

From Eqs. (\ref{gfe}), (\ref{a1}) and (\ref{s1}) it is easy to see that the
only nonanalyticity in the behavior of the free energy exists at $\phi =0.$
Then, for $0<\phi <<1$ one obtains, after some algebra, that the $C$%
-function \cite{remarkd} can be written in the following scaling form 
\begin{equation}
C(t,\lambda )=\frac{\sqrt{\pi /2}}6y_0^{1/4}\exp \left( -\sqrt{y_0}\right) ,
\label{Cd1}
\end{equation}
where the scaling variable is $y_0=\lambda ^2\phi _0/t^2,$ and we have
identified $v=\sqrt{gJ}$. Here 
\begin{equation}
\phi _0=64\exp \left( -4\pi /\lambda \right)   \label{sfed1}
\end{equation}
is the Haldane gap type behavior of the solution of the corresponding
spherical field equation for the zero temperature system. Such type of
solution is very well known from different problems, e. g., one dimensional
anharmonic crystal \cite{PT86} and the quantum nonlinear $O(N)$ sigma model
in the large $N$ limit (see, e.g., \cite{DS93}, \cite{JG94}). In deriving (%
\ref{Cd1}) we have been interested in such a behavior of the nonzero
temperature system which approaches the corresponding zero-temperature
behavior when $T\rightarrow 0.$

As it is clear from the above expressions, $C$ is a positive and a
monotonically increasing function of the temperature.

The behavior of the $C$-function for $d=1$ is illustrated in Fig. 1.

\subsection{ The case $d=2$}

We are interested in the behavior of the $C$-function around and below the
critical point only, i.e. $0<\phi <<1$ \cite{HPT98}, \cite{HDPT97}. As it is
well known the critical point is at $\lambda =\lambda _c=1/{\cal W}_2(0)$ $%
\approx 3.1114$ and $T=0$. Then, taking into account in Eq. (\ref{Cdef})
that $n(2)=\zeta (3)/\left( 2\pi \right) $ and $v=\sqrt{gJ}$, for the
C-function we obtain from Eqs. (\ref{a2}) and (\ref{s2}) that $C(t,\lambda
)=X\left( x\right) ,$ where 
\begin{eqnarray}
X\left( x\right) = && \frac 1{\zeta (3)}\left[ x(y-y_0)+\frac 16\left(
y^{3/2}-y_0^{3/2}\right)\right.\nonumber \\
&  + & \left. \sqrt{y}{\rm Li}_2\left( \exp \left( -\sqrt{y}%
\right) \right) +{\rm Li}_3\left( \exp \left( -\sqrt{y}\right) \right)
\right] ,  \label{Cd3}
\end{eqnarray}
with $x=\pi \left( 1/\lambda -1/\lambda _c\right) \lambda /t.$ Here $%
y=y\left( x\right) $, and $y_0=y_0\left( x\right) $ are solutions of the
corresponding equations that follow from (\ref{Cd3}) by requiring that the
first partial derivative of the r.h.s. of (\ref{Cd3}) with respect to $y$,
and $y_0$, respectively, to be zero. These solutions are 
\begin{equation}
\sqrt{y}=2{\rm arcsh}\left( \frac 12\exp \left( -2x\right) \right) ,
\label{solf}
\end{equation}
and 
\begin{equation}
\sqrt{y_0}=\left\{ 
\begin{tabular}{lll}
$-4x$ & , & $\lambda >\lambda _c$ \\ 
$0$ & , & $\lambda \leq \lambda _c$%
\end{tabular}
\right. .  \label{soli}
\end{equation}
The equation (\ref{Cd3}) determines the {\it exact scaling function of }$C$
for the case $d=2.$ From the above equations one can see the different
behavior of $y$ in the three regions: i) {\it renormalized classical}, where 
$y$ tends to zero exponentially fast as a function of $x$ ($x>>1)$ ii) {\it %
quantum critical}, where $y=O(1)$ (for $x=O(1)$) and iii) {\it quantum
disordered, }where $y$ diverges as $\left( 4x\right) ^2$ for $x<<-1$;{\it \ }
($y\sim \left( \chi t^2\right) ^{-1},$ where $\chi $ is the susceptibility
of the system, see \cite{HPT98}). The location of these regions is depicted
in Fig.~2.

The behavior of the $C$-function reflects the existence of these three
regions.

When $\lambda <\lambda _c$ and $t\rightarrow 0,$ from Eqs. (\ref{Cd3}) -(\ref
{soli}) it follows that 
\begin{equation}
C(t,\lambda )\simeq 1-\frac 1{4\zeta (3)}\exp \left[ -4\pi \left( 1-\lambda
/\lambda _c\right) t^{-1}\right] .  \label{Cmin}
\end{equation}
One explicitly observes the exponentially small corrections to the limit
value of $C=1$ (at $t=0$) that corresponds to massless bosons in $d$
dimensions \cite{CN93}, \cite{S93}.

At $\lambda =\lambda _c,$ $C(t,\lambda )$ simplifies, by using the Sachdev's
identity \cite{S93}, and becomes 
\begin{equation}
C(t,\lambda _c)=\frac 45.  \label{cp}
\end{equation}
This universal {\it rational} number \cite{rem4} has been derived for the
first time for the quantum nonlinear $O(N)$ sigma model in the limit $%
N\rightarrow \infty $ \cite{S93}. It demonstrates that at the quantum
critical point $\lambda =\lambda _c$ the $C$-function does not depend on the
temperature. The difference from the corresponding results in Ref. \cite
{CN93} (compare Fig. 3 in \cite{CN93} with Fig. 3 in this article) is due to
the fact that terms proportional to the difference between $y$ and $y_0$ in (%
\ref{Cd3}) have been neglected there. The above is justified when $y>>1$
(then $y$ and $y_0$ are exponentially close to each other). The analysis of
the corresponding equation shows that the last happens when $x<<-1$ (i.e. $%
\lambda >\lambda _c$) where $y\sim y_0\sim (4x)^2,$ which is the case of the
quantum disordered region. In this case it is easy to see that 
\begin{equation}
C(t,\lambda )\simeq \frac{4\pi }{\zeta (3)}\frac{|1-\lambda /\lambda _c|}t%
\exp \left[ 4\pi \left( 1-\lambda /\lambda _c\right) t^{-1}\right] ,
\label{Cplus}
\end{equation}
i.e. $C$ approaches zero exponentially fast in terms of the scaling
parameter $x.$ The behavior of the $C$-function in this case is that one of
massive free bosons \cite{CN93}.

Let us consider now the monotonicity of the $C$-function. From (\ref{Cd3})
it follows that 
\begin{equation}
\frac{\partial C(t,\lambda )}{\partial t}=-\frac \pi {\zeta (3)}\left(
y-y_0\right) \left( 1-\lambda /\lambda _c\right) t^{-2}.  \label{Cm}
\end{equation}
Since $y>y_0,$ we conclude that $C$ is a monotonically increasing function
of the temperature for $\lambda >\lambda _c,$ and monotonically decreasing
function for $\lambda <\lambda _c.$ Within exponentially small in
temperature corrections this result coincides, in fact, with the
corresponding one for the QNL$\sigma $M in the limit $N\rightarrow \infty $ 
\cite{CN93}.

The above results for the behavior of the $C$-function are illustrated in
Fig. 3.

Finally, it seems worthwhile to mention that, as follows from Eq. (\ref{Cd3}%
), {\it the }$C${\it -function is monotonically increasing function of the
scaling variable }$x${\it \ }(see Fig. 4){\it \ for any value of }$t$ ($%
\lambda /t>>1$).

\subsection{The case $d=4$}

For this case, taking into account that in Eq. (\ref{Cdef}) $n(4)=3\zeta
(5)/\left( 2\pi \right) ^2$ and $v=\sqrt{gJ}$, for the C-function we obtain
from Eqs. (\ref{a4}) and (\ref{s4}) that $C(t,\lambda )=X(x,\lambda /t),$
where 
\begin{eqnarray}
\lefteqn{X(x,\lambda /t) = \frac{\left( 2\pi \right) ^2}{3\zeta (5)}
\left[ x(y-y_0)+%
\frac 14r \left( y^2-y_0^2\right) \frac \lambda t     \right.  }
\label{Cd4} \\
&& +
\left.
\frac 1{\left( 2\pi \right)^2}
\left[
y{\rm Li}_3\left( \exp \left( -\sqrt{y}\right) \right) +3\sqrt{y}%
{\rm Li}_4\left( \exp \left( -\sqrt{y}\right) \right) +3{\rm Li}_5\left(
\exp \left( -\sqrt{y}\right) \right)
\right]
\right]
\nonumber
\end{eqnarray}
with 
\begin{equation}
x=\frac 12\left( \frac \lambda t\right) ^3\left( \frac 1\lambda -\frac 1{%
\lambda _c}\right) ,  \label{x4}
\end{equation}
and $\lambda _c=1/{\cal W}_4(0)\approx $ $5.2882$. Here $y\geq 0$, and $%
y_0\geq 0$ are solutions of the corresponding equations that follow from (%
\ref{Cd4}) by requiring that the first partial derivative of the r.h.s. of (%
\ref{Cd4}) with respect to $y$, and $y_0$, respectively, to be zero. This
leads to the following equation for $y$ 
\begin{equation}
x=-\frac 12r\frac \lambda ty+\frac 1{2\left( 2\pi \right) ^2}\left[ \sqrt{y}%
{\rm Li}_2\left( \exp \left( -\sqrt{y}\right) \right) +{\rm Li}_3\left( \exp
\left( -\sqrt{y}\right) \right) \right] .  \label{eqfd4}
\end{equation}
It is easy to see that, for a given $t$ and $\lambda ,$ the solution of the
above equation, if it exists, is unique. For $y_0$ we get 
\begin{equation}
y_0=\left\{ 
\begin{array}{ll}
-\tilde{x}=-\left( 2t\right) /\left( r\lambda \right) x, & \lambda >\lambda
_c \\ 
0, & \lambda \leq \lambda _c
\end{array}
.\right.  \label{eqid4}
\end{equation}
One observes that in the most general case the function $X,$ given by Eq. (%
\ref{Cd4}), could not be recast in a scaling form. However, as we will see
below, the last is possible in some subregions of the $\lambda -t$ plane. We
recall that the susceptibility $\chi $ of the system is proportional to $%
y^{-1}$ (if $y\neq 0$ \cite{Baxter}, \cite{remark3}), which leads to the
conclusion that a nonzero-temperature phase transition exists at a given $%
t_c=$ $t_c(\lambda ),$ where $t_c(\lambda )$ is given by the equation 
\begin{equation}
t_c(\lambda )=\lambda \left[ \frac{\left( 2\pi \right) ^2}{\zeta \left(
3\right) }\left( \frac 1\lambda -\frac 1{\lambda _c}\right) \right] ^{1/3},
\label{lct}
\end{equation}
(at $t=$ $t_c(\lambda )$ one has $y=0$, and $y=0$ also for $t<t_c(\lambda )$%
). As for the $d=2$ case three principal different regimes exist: i) {\it %
renormalized classical (}where $y$ tends to zero exponentially fast as a
function of $\lambda /t$), ii) {\it quantum critical} (where $y$ tends to
zero algebraically as a function of $\lambda /t$ or $y=O(1)$), and iii) {\it %
quantum disordered (}where $y>>1$). In order to describe the behavior of the 
$C$-function below we analyze these three regimes.

{\it A) }Let{\it \ } us first suppose that $y<<1.$ Then Eq. (\ref{eqfd4})
becomes 
\begin{equation}
\left( \frac \lambda t\right) ^3
\left[ \frac 1\lambda -\frac 1{\lambda _c}-%
\frac{\zeta (3)}{\left( 2\pi \right) ^2}\left( \frac t\lambda \right)^3
\right]
=\frac 1{\left( 4\pi \right) ^2}y\ln \left( y/e\right) -r\frac %
\lambda ty.  \label{eqsy}
\end{equation}
Obviously, there are two subregimes {\it a)} when the first term in the
right hand side dominates and {\it b)} when the second one dominates. The
borderline between them is given by 
\begin{equation}
\frac 1{2r}\left( \frac \lambda t\right) ^2
\left[ \frac 1{\lambda _c}-\frac 1%
\lambda +\frac{\zeta (3)}{\left( 2\pi \right) ^2}\left( \frac t\lambda
\right) ^3
\right] =\exp \left[ -\left( 4\pi \right) ^2r\frac \lambda t%
+1\right] .  \label{linelog}
\end{equation}
In the $\lambda -t$ plane Eq. (\ref{linelog}) determines a line $%
t^{*}(\lambda )$ that is exponentially close to (as a function of $\lambda /t
$ ) the line $t_c\left( \lambda \right) .$  At $t^{*}(\lambda )$ the
solution of Eq. (\ref{eqsy}) is 
\begin{equation}
y\sim \exp \left[ -\left( 4\pi \right) ^2r\frac \lambda t\right] ,
\label{atlstar}
\end{equation}
whereas $y=0$ at $t_c(\lambda ).$ We conclude that the {\it renormalized
classical regime} is observed for parameters of the system laying in the $%
\lambda -t$ plane between $t_c\left( \lambda \right) $ and $t^{*}(\lambda ).$
In this regime one could neglect the second term in the r.h.s. of (\ref{Cd4}%
) which leads to a scaling form of the $C$-function with a scaling variable $%
x,$ defined in (\ref{x4}). At $t^{*}(\lambda )$ the $C$-function could not
be rewritten in a scaling form. In the remainder we will see that another
scaling variable $\tilde{x}$ can be defined for the region to the right of $%
t^{*}(\lambda ).$ This is due to the fact that in this region the first term
in the r.h.s. of Eq. (\ref{eqfd4}) is of the leading order. Indeed, this is
true not only for the case {\it b) }but also for the cases {\it B)}, when $%
y=O(1)$, and {\it C),} when $y\gg 1,$ which cases are to be considered below.

Before passing to the consideration of case {\it B)} let us note that the
further inspection of Eq. (\ref{eqsy}) for the case {\it b)} leads to the
conclusion that there exists a crossover line 
\begin{equation}
t_s(\lambda )=\lambda \left[ \frac{\left( 2\pi \right) ^2}{\zeta \left(
3\right) }\left( \frac 1{\lambda _c}-\frac 1\lambda \right) \right] ^{1/3},
\label{lplus}
\end{equation}
between two regimes where $\chi (t,\lambda )\sim t^{-3}$ and $\chi
(t,\lambda )\sim t^{-2}$, respectively. This curve is symmetric to the curve 
$t_c(\lambda )$ with respect to the line $\lambda =\lambda _c.$ Let us turn
now to the case

{\it B) }$y=O(1).$ Then, since $t<<1,$ Eq. (\ref{eqfd4}) becomes extremely
simple and, up to the leading order coincides with the corresponding
equation for the zero-temperature system (see Eq. (\ref{eqid4})). Its
solution is 
\begin{equation}
y=\frac 1r\left( \frac \lambda t\right) ^2\left( \frac 1{\lambda _c}-\frac 1%
\lambda \right) \equiv -\tilde{x}.  \label{solyo1}
\end{equation}
Since $\chi (t,\lambda )\sim \left( yt^2\right) ^{-1}$ and $y=O(1),$ one
concludes that in this regime $\chi (t,\lambda )\sim t^{-2}$. In a given
sense a formal curve $t_1(\lambda )$ in the $\lambda -t$ plane which borders
the region in which $y=O(1)$ can be obtained by simply setting $y=1$ in Eq. (%
\ref{solyo1}). Summarizing the results from {\it A)} and {\it B)} we are
lead to the conclusion that the {\it quantum critical regime} is observed
for values of the parameters $t$ and $\lambda $ laying in the $\lambda -t$
plane between the curves $t^{*}(\lambda )$ and $t_1(\lambda ).$ We see also
that in this region $X(t,\lambda )=X(\tilde{x}),$ i.e. the scaling is
restored with the new scaling variable $\tilde{x}.$ We pass now to the case

C) $y>>1.$ Then one formally receives the same solution as given by Eq. (\ref
{solyo1}) but now $\chi (t,\lambda )\sim \left( 1/\lambda _c-1/\lambda
\right) ^{-1},$ i.e. it does not depend on $t$ up to exponentially small in $%
\lambda /t$ corrections. We conclude that the region of parameters in the $%
\lambda -t$ plane below $t_1(\lambda )$ determines the {\it quantum
disordered }region. Again, as in B), the scaling variable of $C$ is $\tilde{x%
}$.

The above results are summarized in Fig. 5.

The existence of the regions of thermodynamic parameters determined above is
reflected by the corresponding behavior of the $C$-function given by Eq. (%
\ref{Cd4}).

First, for $t\leq t_c(\lambda )$ (i.e. under the existence of a long-range
order in the system), since $y=y_0=0,$ one immediately obtains from (\ref
{Cd4}) that $C=1$ \cite{remanyd}.

Further, for $t>t_c(\lambda )$ taking into account Eqs. (\ref{eqid4}) and (%
\ref{eqfd4}), it is easy to see that 
\begin{equation}
\frac{\partial C(t,\lambda )}{\partial t}=-\frac{2\pi ^2r}{\zeta (5)}\frac %
\lambda {t^2}\left( y-y_0\right) \left[ \tilde{x}+\frac 16r\left(
y+y_0\right) \right] .  \label{derc4}
\end{equation}
Since $y>y_0>0$ for the considered region of parameters $\lambda $ and $t$,
the above equation leads us to the conclusion that for any $\lambda <\lambda
_c$ the $C$-function is a {\it monotonically decreasing function of the
temperature}. The same is true also for $\lambda =\lambda _c$ and $t\neq 0.$
From the analysis of the solutions of the equations for $y$ and $y_0$ it
becomes clear that for a fixed $\lambda $ and $t>t_s(\lambda )$ the {\it %
leading-order} form for both $y$ and $y_0$ is $y=y_0=-\tilde{x}$ (if one
takes into account next-to-the-leading order terms then, of course, $y>y_0$%
). Setting the above expressions for $y$ and $y_0$ in the rectangular
brackets in Eq. (\ref{derc4}), we conclude that $C$ is a {\it monotonically
increasing function of }$t${\it \ }for any $t>t_s(\lambda ).$ It is clear
that somewhere between $\lambda =\lambda _c$ and $t_s(\lambda )$ the
derivative of the $C$-function changes its sign, i.e. there is a line $%
t_{st}(\lambda )$ of stationary points $\partial C(t,\lambda )/\partial t=0$
. One can see that 
\begin{equation}
t_{st}(\lambda )=\lambda \left[ \frac{\left( 4\pi \right) ^2}{\zeta \left(
3\right) }\left( \frac 1{\lambda _c}-\frac 1\lambda \right) \right] ^{1/3}.
\label{lst}
\end{equation}
Since the point $\lambda =\lambda _c,t=0$ lies on $t_{st}(\lambda )$ and at
it $C=1,$ we conclude that $C=1$ at the whole line $t_{st}(\lambda ).$ It is
clear now that $C$ is a {\it monotonically increasing} function of $t$ for $%
t<t_{st}(\lambda ).$ It is a {\it monotonically decreasing} function of $t$
for $t>t_{st}(\lambda )$ as well as for any (small) $t$ if $\lambda <\lambda
_c.$ These results are summarized in Fig. 6.

\section{Finite-size scaling interpretation}

It is interesting to interpret the bulk critical behavior of the $C$%
-function in the context of the finite-size scaling (FSS) theory by
introducing a finite ``temporal'' dimension $L_\tau =\lambda /t.$ Then,
taking into account: {\it i)} the dimensional crossover rule that connects
the properties of a given $d$-dimensional quantum system and those ones of
the corresponding $(d+1)$-dimensional classical one
with the mapping
$d \rightarrow \d+1, L \rightarrow L_\tau, t\rightarrow \lambda $,
{\it ii)} the
Privman and Fisher \cite{FP85} hypothesis for the free energy of a finite
classical system when the hyperscaling holds (i.e. between the lower $d_l$
and the upper $d_u$ critical dimensions of the system),
one could make the statement that the free energy
$f_\infty $ of a quantum system with
dimensionality $d_l<d<d_u$ should have the form

\begin{equation}
\left[f_\infty \left( t,\lambda |d\right) -f_\infty \left( 0,\lambda
|d\right)\right]/(T L_\tau)^{-1}
= L_\tau ^{-(d+1)}Y\left( L_\tau /\xi \left(
0,\lambda \right) \right) ,  \label{h1}
\end{equation}
where $\xi \left( 0,\lambda \right) $ is the correlation length of the
zero-temperature system, and $Y$ is a {\it universal function. }We remind
that{\it \ }in order to have no{\it \ }nonuniversal prefactor in front of $Y$
for classical systems one considers $\tilde{f}_\infty =\beta f_\infty ,$
instead of $f_\infty $ itself. The normalization of the free energy in (\ref
{h1}) simply follows by our choice of $L_\tau .$ For the
model considered here the inspection of Eq. (\ref{Cd3}) shows that the
hypothesis (\ref{h1})
is indeed valid with the standard scaling variable $L_\tau /\xi
\left( 0,\lambda \right) \equiv x=\pi \left( 1/\lambda -1/\lambda _c\right)
\lambda /t$ and $C\left( t,\lambda \right) $ $=X(x)$ $=-Y(x)/n(d).$ It is
interesting that despite the lack of hyperscaling at $d_l=1$ the $C$%
-function again can be written as a function of $L_\tau /\xi \left(
0,\lambda \right) $ (see Eq. (\ref{Cd1})), if one identifies $\xi \left(
0,\lambda \right) =\phi _0^{-1/2},$where $\phi _0$ is given by Eq. (\ref
{sfed1}). The case $d=4>d_u$  is much more interesting due to the lack of
hyperscaling. In the most general case the $C$-function could not be even
recast in a FSS form (see Eq.(\ref{Cd4})). The last is possible
exponentially close (in $L_\tau )$ to the line of finite-temperature phase
transitions $t_c\left( \lambda \right) ,$where the modified scaling variable
is $2x=L_\tau \left[ L_\tau /\xi \left( 0,\lambda \right) \right] ^2$ (see
Eq. (\ref{x4})). The standard scaling variable $\tilde{x}=L_\tau /\xi \left(
0,\lambda \right) $ is restored only for parameters to the right of the
curve $t^{*}\left( \lambda \right) $ in the $\lambda -t$ plane (see Eq. (\ref
{solyo1}) and the comments connected with it). This change of scaling
variables from $x$ to $\tilde{x}$ is a new point within FSS theory. Normally
one observes modified FSS (see \cite{Preview}, \cite{BTreview}) above $d_u$
due to the existence of a dangerous irrelevant variables in the system. On
the other hand, considering the $d=5$ dimensional spherical model film
Barber and Fisher, as early as in 1973 \cite{FB73}, stated that the scaling
variable should be the standard one, i.e. $L_\tau /\xi \left( 0,\lambda
\right) $ in our notations. The above results resolve this seeming
contradiction: the scaling variable has to be modified very close to the
phase boundary, but is the standard one a bit away of it. The physical
reasoning for that difference is the existence in the system of a
temperature driven phase transition in addition to the quantum one with
respect to $\lambda $ at $t=0.$ To our knowledge all other examples
considered previously in the literature of modified FSS concerns finite
systems with no (sharp) phase transition in it.

\section{Concluding remarks}

One generally expects that  the $C$-function increases  monotonically when
the quantum fluctuations ``dominate'' \cite{CN93}. The real meaning of the
term ``dominate'' turns out to be quite subtle, as we have demonstrated in
the current article. In fact, we have shown that the region where the $C$%
-function remains monotonically increasing (as a function of temperature)
and the quantum critical region do essentially intersect but do {\it not}
coincide (see Fig. 2). This is one of the results of the present work. The
question of where one should look for and what should be understood as
domination of quantum fluctuations is, indeed, very intriguing. It is a part
of the more general problem of a {\it quantitative} description of the
interplay of the quantum and critical fluctuations. There exist different
views on that issue. The standard one \cite{Sreview}, \cite{CJT89} is based
on the ``ratio'' between the correlation length and the length of De
Brougle. Another possible approach can be based on the behavior of the $C$%
-function \cite{CN93}, \cite{Z97}. Furthermore, there is an approach based
on the algebra of critical fluctuation operators, due to Verbeure and
Zagrebnov \cite{VZ92}, where a measure of the ``degree of criticality'' is
introduced in a mathematically rigorous way.

In the present work we investigated the behavior of the $C$-function for $%
d=1,2,$ $4.$ The case $d=1$ represents the situation with no phase
transition and strong quantum fluctuations, $d=2$ -- the one when a quantum
critical point appears at $T=0,$ and $d=4$ -- when there is a line of
classical critical points ending up with a zero temperature (quantum)
critical point. In fact, these are the most typical cases on which the
attention in the literature is focused.

{\it Case} $d=1.$ As it is to be expected on general grounds, the $C$%
-function increases monotonically as a function of temperature (see Fig. 1).
This reflects the fact that the quantum fluctuations are strong enough (as
it is clear from Eq. (\ref{sfed1}) one cannot consider $\lambda $ as a small
parameter) and the lack of a critical point. The $C$-function obtained here
coincides with the $C$-function of the massive free bosons (for $d=1$) with
mass $\sqrt{\phi _0}$ , because of the exponentially small difference then
between $\phi $ and $\phi _0$, i.e. one can consider $\phi $ as a fixed
parameter in (\ref{a1}) and in (\ref{s1}). The general case (for any $d$) of
free massive bosons actually follows from (\ref{gfe}) - (\ref{R}) by
considering $\phi $ there as a fixed parameter connected to the mass $m$ of
bosons ($\phi \sim m^2$ ). For the last case it is trivial to check that the
corresponding $C$-function is that one obtained in \cite{CN93} (see Eq.
(3.2) there).

{\it Case} $d=2.$ Fig. 2 shows the phase diagram for our model which
coincides with the phase diagram of the $d=1$ quantum Ising model, as well
as with the nonlinear $O(n)$ sigma model in the limit $n\rightarrow \infty $%
, see, e.g., Sachdev \cite{Sreview}. As a function of the temperature $C$ is
monotonically increasing for $\lambda $ above $\lambda _c,$ equals $4/5$ at $%
\lambda =\lambda _c$ (and then $C$ does not depend on $t$) and is
monotonically decreasing for $\lambda $ below $\lambda _c$ (see Fig. 3). The
lack of overall monotonicity with respect to the temperature is due to the
crossover from classical to quantum behavior. It is clear, that one indeed
can consider {\it monotonicity of }$C${\it \ as a measure} of the role the
corresponding fluctuations are playing in a given region of parameters. It
is interesting that $C$ changes its monotonicity, in fact, in the {\it middle%
} of the quantum critical region. Finally, we note that it is nevertheless
possible to find a (nontrivial) variable, with respect to which the $C$%
-function is monotonic in the whole $t-$ $\lambda $ plane (see Fig. 4). This
variable is the scaling variable $x=\pi \left( 1/\lambda -1/\lambda
_c\right) \lambda /t$.

{\it Case} $d=4.$ The existence of a line of non-zero temperature critical
points modifies drastically the corresponding picture in comparison with the 
$d=2$ case. Now a line of stationary points $t_{st}(\lambda )$ appears (see
Fig. 5) which ``starts'' from $(\lambda =\lambda _c,t=0)$ and lies to the
left of $t_s(\lambda ).$ To the left of $t_{st}(\lambda ),$ $C$ is a
non-increasing function of the temperature (see Fig. 6). For $\lambda
<\lambda _c$ and $t<t_c(\lambda )\ $one has $C=1,$ whereas within the region
between $t_c(\lambda )$ and $t_{st}(\lambda )$ the $C$-function is
monotonically decreasing as a function of the temperature. To the right of $%
t_{st}(\lambda )$ the $C$-function becomes monotonically increasing function
of the temperature, being zero at the $t=0,$ $\lambda >\lambda _c$ line. At
the lines $t_c(\lambda )$ and $t_{st}(\lambda )$ the $C$-function reaches
its maximum value, i.e. it becomes $C=1.$ Finally, we would like to mention
that, similar to the case $d=2,$ it is possible to find two nontrivial
parameters such that with respect to both of them the $C$-function is
monotonically increasing. Such are, e.g., the parameters $x$ and $\lambda /t$
(see Eq. (\ref{Cd4}) and take into account that $y>y_0$).

Comparing the behavior of the $C$-function for $d=1,$ $d=2$ and $d=4$ we
conclude that

a) For $d=1$ for any fixed $\lambda $ we have a monotonically increasing
with temperature $C$-function.

b) For $d=2$ the above is true only for $\lambda >\lambda _c.$

c) For $d=4$ the $C$-function is a monotonically increasing function of $t$
for $\lambda >\lambda _c$ and $t$ {\it small enough}. The monotonicity of
the $C$-function does not change by increasing $t$ {\it only} for $\lambda
=\lambda _c.$

So, the region in the parametric space where $C$ remains monotonically
increasing with $t$ becomes smaller when $d$ increases. We explicitly see
the crucial role the dimensionality $d$ and the existence of phase
transition, which appears upon increasing $d,$ play in the behavior of the $%
C $-function as a function of $t.$ Nevertheless, for any $d$ one can find
nontrivial variable(s), function(s) of the temperature and the parameter
controlling the quantum fluctuations, in terms of which $C$ is a
monotonically increasing function of its variable(s). In close vicinity of
the quantum critical point the $C$-function is given by a universal scaling
function which properties can be interpreted in terms of FSS that has to be
modified for $d=4.$

\section*{Acknowledgments}

This work is partially supported by Bulgarian National Science Foundation
under grants F-608/96 and MM-603/96. The authors thanks J. G. Brankov for
valuable discussions on this work.

\newpage

\begin{center}
{\bf Figure Captions}
\end{center}

Fig. 1. The bold line $t_{LT}=8\lambda \exp \left( -2\pi /\lambda
\right) $ borders from above the region in the $t-$ $\lambda $ plane where
the expression for the $1$-d $C$-function, given by Eq. (\ref{Cd1}), is
valid. The symbol $C\uparrow $ means that $C$ increases in whole that region
starting form $C=0$ at $t=0$.

 Fig. 2. The phase diagram of the model and crossovers for the case
$d=2$ as a function of $t$ and the normalized quantum parameter $\lambda .$
One distinguishes renormalized classical, quantum critical and quantum
disordered regions. Long-range order is present only at $t=0$ for $\lambda
<\lambda _c.$

 Fig. 3. The behavior of the $2$-d $C$-function is illustrated for $%
\lambda =1.5\lambda _c,$ $\lambda =\lambda _c$ and $\lambda =0.5\lambda _c.$

 Fig. 4. The behavior of $2$-d $C$ as a function of the scaling
parameter $x=\pi \left( 1/\lambda -1/\lambda _c\right) \lambda /t$ .

 Fig. 5. The phase diagram of the model and crossovers for the case 
$d=4.$ Long-range order exists below the line $t_c.$ The line $t_{st}$ is
the locus of points in the thermodynamic space where $\partial C/\partial
t=0.$ The other lines denote crossovers between different regimes which are
described in the text.

 Fig. 6. The monotonic behavior of $C$ as a function of temperature
is shown. The symbols $C\uparrow $ ( $C\downarrow $) mean that $C$ is a
monotonically increasing (decreasing) function of the temperature. The
number in ellipses show the value of $C$ at the corresponding line. In whole
long-range order region, i.e. below the line $t_c,$ $C=1.$


\section*{References}
\begin{thebibliography}{99}


\bibitem{Z86}  Zamolodchikov A B 1986 {\it JETF Lett.} {\bf 43} 731;
1987 {\it Sov. J. Nucl. Phys.} {\bf 46} 1090.

\bibitem{J91}  Jack I 1992, {\it Renormailzation Group 91}, edited by D. V.
Shirkov and V. B. Priezzhev (World Scientific, Singapore, 1992), pp. 55 -
65.

\bibitem{CN93}  Castro Neto A H and Fradkin E 1993
{\it Nuclear Physics B} {\bf 400}[FS] 525.

\bibitem{S93}  Sachdev S 1993 {\it Phys. Lett. B} {\bf 309} 285.

\bibitem{Z97}  Zabzin M, hep-th/9705015.

\bibitem{PV98}  Petkou A C and Vlachos N D hep-th/9803149.

\bibitem{Cardy87}  Cardy J L 1987 {\it Nucl. Phys. B} {\bf 290} 355.

\bibitem{V96}  Vojta T 1996 {\it Phys. Rev. B} {\bf 53} 710.

\bibitem{rem2}  The lattice constant is taken to be $a=1$ and the dependence
on it will be omitted hereafter.

\bibitem{HPT98}  Chamati H, Pisanova E S and Tonchev N S 1998
{\it Phys. Rev. B} {\bf 57} 5798.

\bibitem{HDPT97}  Chamati H, Danchev D M, Pisanova E S and Tonchev N S,
cond-mat/9707280 ICTP, Trieste, preprint IC/97/82, July 1997.

\bibitem{PT86}  Plakida N M and Tonchev N S 1986
{\it Physica A} {\bf 136} 176.

\bibitem{DS93}  S\'{e}n\'{e}chal D 1993
{\it Phys. Rev. B} {\bf 47} 8353.

\bibitem{JG94}  Jolicoeur Th and Golinelli O 1994
{\it Phys. Rev. B} {\bf 50} 9265.

\bibitem{remarkd}  For the simplicity of the notations the dependence on the
argument $d$ of the $C$-function is omitted hereafter.

\bibitem{rem4}  Let us note that $\zeta (3)$ is irrational number, as was
pointed out by Ap\'{e}ry, 1978 (see \cite{J90}). Therefore, none of the
intermediate steps suggests that a rational number will be the final result.

\bibitem{J90}  Jullia B 1990 {\it Number theory and physics, }Springer
Proceedings in Physics {\bf 47}, edited by J. M. Luck, P. Moussa and M.
Waldschmidt (Springer, Berlin, 1990), p. 276.

\bibitem{Baxter}  Baxter R J 1982
{\it Exactly solved models in statistical
mechanics,} (Academic Press, London, 1982).

\bibitem{remark3}  If $y=0$ the relation between the susceptibility and $y$
is a bit more subtle for dimensionalities above the upper critical
dimension; see, e.g. \cite{Baxter}, Chapter 5.

\bibitem{remanyd}  Note that this result does depend only on the existence
of long-range order in the system (then $y=y_0=0$) and not on the
dimensionality $d.$ From Eqs. (\ref{Cdef}) and (\ref{gfe}) - (\ref{ftt}) and
the identity 
\[
\Gamma \left( \frac s2\right) \pi ^{-s/2}\zeta \left( s\right)
=\int_0^\infty R\left( \pi s\right) x^{s/2}\frac{dx}x,\qquad 
%TCIMACRO{\func{Re}}
%BeginExpansion
\mathop{\rm Re}%
%EndExpansion
s>1,
\]
we get $C=1.$

\bibitem{Sreview}  Sachdev S 1996 {\it {Strongly Correlated Magnetic and
Superconducting Systems}}, edited by G. Sierra and M. A. Martin-Delgado
(Springer, Berlin, 1996); Sondhi S L, S. M. Girvin, Carini J P and 
Shahar D 1997 {\it Rev. Mod. Phys.} {\bf {69}} 315.

\bibitem{CJT89}  Continentino M A, Japiassu G M and Troper A 1989
{\it Phys. Rev. B} {\bf 39} 9734; Sengupta A M and Georges A 1995
{\it Phys. Rev. B} {\bf 52} 10 295;
Millis A J 1993 {\it Phys. Rev. B} {\bf 48} 7183;
Sachdev S, Read N and Oppermann R 1995
{\it Phys. Rev. B} {\bf 52}, 10 286.

\bibitem{VZ92}  Verbeure A and Zagrebnov V A 1992
{\it J. Stat. Phys.} {\bf 69} 329;
Verbeure A and Zagrebnov V A 1995 {\it J. Stat. Phys.} {\bf 79} 377;
Momont B, Verbeure A and Zagrebnov V A 1997
{\it J. Stat. Phys.} {\bf 89} 633.

\bibitem{FP85}  Privman V and Fisher M E 1984
{\it Phys. Rev. B} {\bf 30} 322.

\bibitem{Preview}  Privman V 1990 {\it {Finite Size Scaling and Numerical
Simulations of Statistical Systems}}, edited by V. Privman (World
Scientific, Singapore, 1990).

\bibitem{BTreview}  Brankov J and Tonchev N 1992
{\it Physica A} {\bf 189} 583.

\bibitem{FB73}  Barber M N and Fisher M E 1973
{\it Ann. Phys.} {\bf 77} 1.
\end{thebibliography}
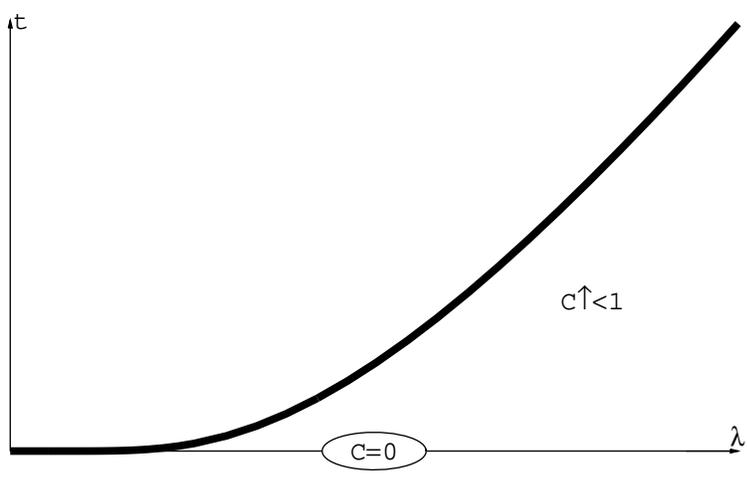
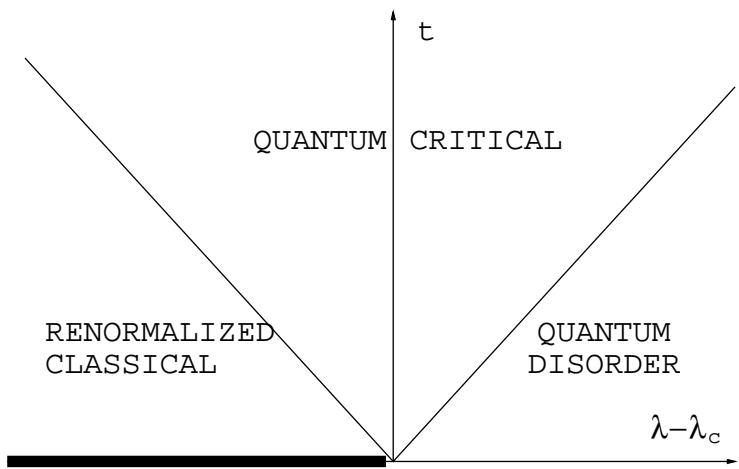
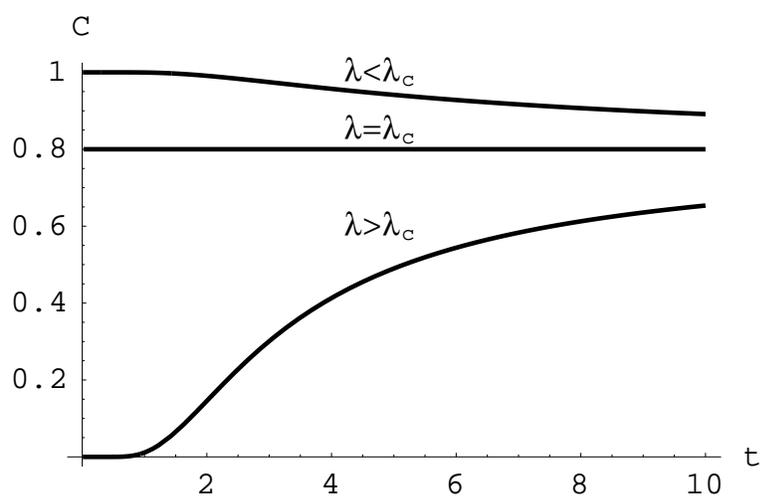
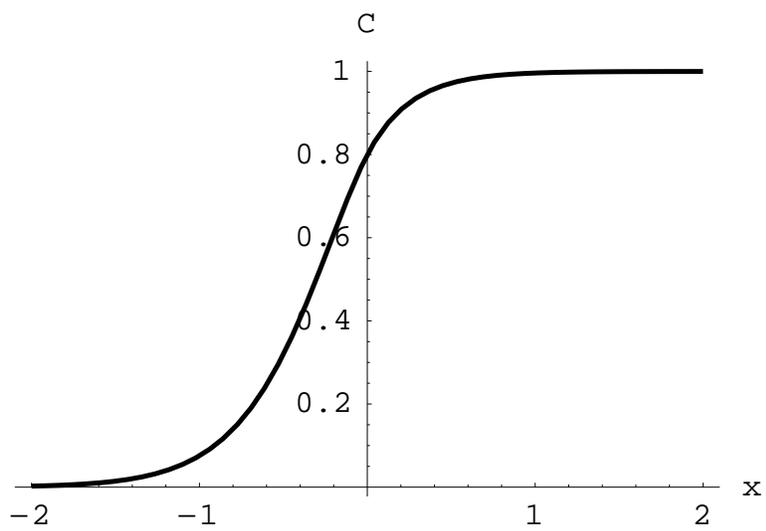
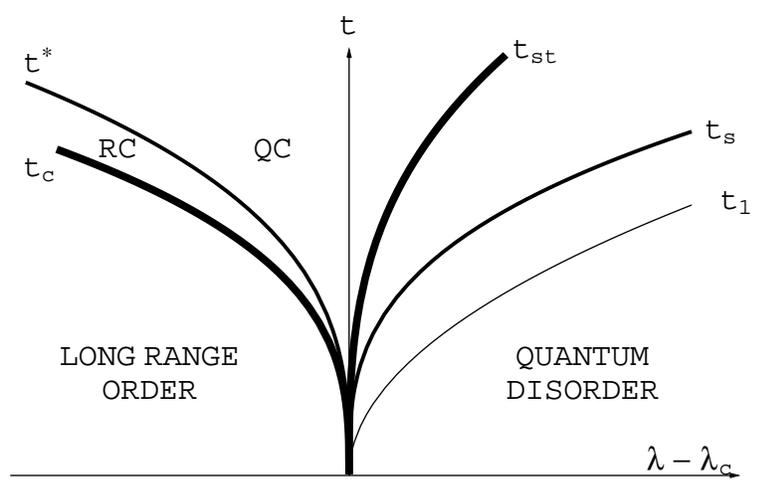
\end{document}